\documentclass[aps,pra,twocolumn,superscriptaddress,longbibliography,nofootinbib,showkeys,showpacs]{revtex4-1}
\usepackage{blkarray}
\usepackage{amsfonts}
\usepackage{amsmath}
\usepackage{graphicx}
\usepackage{subcaption}
\usepackage[justification=raggedright]{caption}  
\usepackage{appendix}
\usepackage{bm}
\usepackage{float}
\usepackage{braket}
\usepackage{url}
\usepackage{xcolor}
\usepackage{comment}
\usepackage[top=2cm,left=2cm,right=2cm,bottom=2cm]{geometry}
\usepackage{hyperref}
\usepackage[utf8]{inputenc}
\usepackage{CJK}
\usepackage[capitalise]{cleveref}
\usepackage{cases}
\usepackage[ruled,vlined,linesnumbered,boxed]{algorithm2e}
\SetKwFunction{ESOP}{ESOP}
\SetKwInOut{Input}{Input}
\SetKwInOut{Output}{Output}

\begin{document}
\colorlet{CV}{.}
\SetKwRepeat{Do}{do}{while}
\title{Variational Quantum Algorithm for Constrained Combinatorial Optimization Problems}

\author{Hui-Min Li}
\email{lihuimin199301@163.com}
\affiliation{College of Science, North China Institute of Science and Technology, Langfang, Hebei 065201, China}
\author{Yuan-Liang Han}
\affiliation{College of Science, North China Institute of Science and Technology, Langfang, Hebei 065201, China}
\author{Zhi-Xi Wang}
\email{wangzhx@cnu.edu.cn}
\affiliation{School of Mathematical Sciences, Capital Normal University, 100048, Beijing, China}
\author{Shao-Ming Fei}
\email{feishm@cnu.edu.cn}
\affiliation{School of Mathematical Sciences, Capital Normal University, 100048, Beijing, China}

\begin{abstract}
While variational quantum algorithms (VQAs) have demonstrated considerable success in unconstrained optimization, their application to constrained combinatorial problems face a trade-off. Penalty-based methods, despite their circuit simplicity, suffer from a fundamental limitation: inefficient sampling in vast infeasible regions. This often results in suboptimal solutions that violate constraints and impede convergence to high-quality results. In contrast, ansatz-based approaches enforce solution feasibility by design but require complex, problem-specific circuits that are challenging to implement on current noisy intermediate-scale quantum devices. To overcome these limitations, we introduce an alternative VQA whose core innovation lies in a strategically designed loss function. This function offers a dual advantage. First, it is provably guaranteed that its global minimum corresponds uniquely to the optimal feasible solution, as this is achieved by ensuring universally higher loss values for all infeasible solutions. Second, it furnishes distinct computational pathways for feasible versus infeasible regions, thus creating clear and non competing guidance for the optimizer. As a result of these combined features, the algorithm's overall performance is significantly enhanced. Regarding hardware overhead, our design requires adding only an efficient validation oracle module to the penalty-based circuit, resulting in a circuit complexity significantly lower than that of ansatz-based approaches with their custom mixers. To validate the practical efficiency of our method, we empirically demonstrate its effectiveness by solving minimum vertex cover and maximum independent set problems on random graphs of varying small-scale sizes.

\end{abstract}
\maketitle

\section{Introduction}
Combinatorial optimization aims to find an optimal object in discrete solution space, and has various applications in science and industry such as supply chain optimization \cite{hao_quantum-inspired_2022,ESKANDARPOUR201511}, transportation networks \cite{Yao2017}, vehicle routing and portfolio investment \cite{Juarna2017,Sbihi2007,zhou2023}. However, the intrinsic hardness of most combinatorial problems renders these optimization tasks computationally challenging on classical computers \cite{Shor1994}. Driven by its computation complexity and fundamental importance, there is immense interest in solving combinatorial optimization problems by using quantum algorithms such as quantum adiabatic algorithms \cite{RevModPhys.80.1061,Farhi2001AQA,mehta_quantum_2021}, digitized adiabatic quantum computation \cite{Barends2016DigitizedAQ,chandarana_digitized-counterdiabatic_2022} and
variational quantum algorithms (VQAs)\cite{cerezo_variational_2021,PhysRevResearch.4.033142}.

As hybrid quantum-classical algorithms, VQAs have emerged as the leading strategy to attain quantum advantages on noisy intermediate-scale quantum devices \cite{cerezo_variational_2021,Preskill_2018,Peruzzo_2014}. Since each gate operation involves inevitable noise and a VQA requires an outer loop classical optimization to assign the variational parameters, leveraging a VQA with a shallow depth becomes paramount for achieving optimal performance.

As one popular VQA, the quantum approximate optimization algorithm (QAOA) was originally proposed by Farhi $\emph{et al.}$ \cite{farhi_quantum_2014}, aiming to find the approximate solutions in combinatorial optimization problems. In a $p$-depth QAOA, the approximate solution to a combinatorial optimization problem is encoded in a wave-function ansatz
\begin{equation}\label{eq:QAOA_ansatz}
\ket{\vec{\gamma},\vec{\beta}}=e^{-i\beta_p\hat{B}}e^{-i\gamma_p\hat{C}}\,\cdots\,
e^{-i\beta_1\hat{B}}e^{-i\gamma_1\hat{C}}\ket{s},
\end{equation}
where $\hat{B}$ is a mixing Hamiltonian, $\hat{C}$ is a diagonal problem Hamiltonian whose ground state encodes the solution, and $\ket{s}$ is some initial state. While QAOA and other quantum algorithms show promise for addressing combinatorial optimization problems with potential computational benefits over classical methods \cite{zhou2023,wurtz_maxcut_2021,crooks_performance_2018,Guerreschi2019,Lucas2014,ruan_quantum_2023}, they are still challengeable when tackling constrained combinatorial optimization problems that are prevalent in practical applications.

Recently, researchers have devised several quantum algorithmic approaches to tackle constrained combinatorial optimization problems. One method involves ansatz-based schemes \cite{Hadfield2018,Hadfield2019,Hadfield2017,Saleem2021,Marsh2019,Marsh2020,Progressive_quantum_algorithm}.
By encoding constraints into $\hat{B}$ and preparing an initial feasible solution $\ket{s}$, these schemes ensure that the mixing operation occurs only within the feasible solution space, thereby guaranteeing the generation of only feasible outputs. While the ansatz-based schemes ensure only feasible outputs, their implementation is hindered by the intricate construction of $\hat{B}$, which necessitates deeper quantum circuits at each computational step. An alternative approach encodes the constrained combinatorial optimization problem into the Ising Hamiltonian by appending so-called penalty terms to the problem Hamiltonian $\hat{C}$ upon constraint violations, which effectively converts the constrained problem into an unconstrained one for simplified resolution \cite{Lucas2014,lihuimin2023Ising}. Nevertheless, although this approach offers shallower quantum circuits compared to the ansatz-based methods, it suffers from inefficient sampling in infeasible regions, which hinders convergence to high-quality solutions and may lead to suboptimal outcomes that may violate the constraints. Furthermore, this approach incorporates a penalty term which introduces additional hyperparameters, and whose selection can significantly impact the algorithm's performance.

This work presents an alternative VQA targeting non-deterministic polynomial-time optimization (NPO) problems (as defined in \cite{Marsh2019,kann1995}), a class of constrained combinatorial optimization problems for which it is efficient to determine whether a solution is feasible. The core innovation of our approach lies in a strategically designed loss function, rather than mapping the problem to one Hamiltonian. This design not only eliminates the need for penalty terms but also obviates the construction of complex mixing operators. The algorithm implementation involves two key steps.

\begin{enumerate}
    \item \textbf{Feasibility certification via ancilla qubit} \\
      We employ an ancilla qubit as a quantum feasibility flag, entangled with the parametrized work qubits through a validation oracle $\hat{U}_v$. This hybrid encoding ensures that measuring the ancilla in state $|1\rangle$ directly certifies the feasibility of the corresponding solution encoded in the work qubits, while $|0\rangle$ indicates infeasibility.

    \item \textbf{A Feasibility-guiding loss function} \\
    Leveraging ancilla-based feasibility certification, we design a loss function that assigns universally higher values to infeasible solutions, guaranteeing that its global minimum corresponds uniquely to the optimal feasible solution, and establishes distinct computational pathways for feasible versus infeasible regions, thus delivering clear, non-competing guidance to the optimizer. Theoretical analysis and detailed construction are provided in Sec.\ref{sec:VQAs}.

\end{enumerate}

Our algorithmic framework is grounded on the assumption of an efficient validation oracle $\hat{U}_v$, where efficiency is defined as the polynomial scaling of circuit resources, including gate count, depth, and number of ancilla qubits, with the problem size. This assumption is justified for NPO problems, as verifying solution feasibility is classically efficient, guaranteeing the existence of an efficient classical circuit. This classical circuit can be transformed into an efficient quantum circuit via reversible logic. For practical implementation, the corresponding oracle $\hat{U}_v$ is constructed using the exclusive-sum-of-products (ESOP) formulation of the constraints, as this approach leads to compact quantum circuits \cite{Marsh2019,Stergiou2004}.

The polynomial-size quantum circuit ensures that the corresponding ESOP representation has a number of product terms scaling polynomially with the problem size. This follows because the circuit consists of a polynomial number of reversible gates, each admitting an ESOP expression with a fixed number of terms \cite{Marsh2019,Drechsler1999,Bernasconi2023,Zulehner2019}. In practice, heuristic ESOP minimization techniques \cite{Stergiou2004,Mishchenko2001,Sasao1993,song1996,Stergiou2002} are applied to find a near-optimal expression, directly reducing the quantum circuit's complexity and enhancing implementation efficiency.

The paper is structured as follows. In Sec. \ref{sec:ESOP} we provide a concise overview of ESOP formulation, along with an analysis of the quantum resource complexity associated with the circuit implementation of each product term. In Sec. \ref{sec:VQAs} we begin by reviewing the limitations of penalty-based methods and the high circuit complexity of ansatz-based schemes. In contrast, we introduce our VQA, comprising two key components: a parametrized ansatz employing an ancilla-based quantum feasibility flag and a strategically designed loss function that provides distinct guidance for optimization across feasible versus infeasible regions. In Sec. \ref{sec:Experiments} we benchmark the performance of our approach against the penalty-based method described in \cite{lihuimin2023Ising} on the minimum-vertex-cover (MVC) and maximum-independent-set (MIS) problems. A summary is given and results are discussed in Sec. \ref{conclusion}.

\section{validation oracle $\hat{U}_v$ based on ESOP expressions}
\label{sec:ESOP}
The constraints of an NPO problem can be characterized by a boolean function $f: \{0,1\}^n \to \{0,1\}$ \cite{Marsh2019,ruan_quantum_2023}. Here, an input vector $(x_1, \ldots, x_n) \in \{0,1\}^n$ represents a candidate solution, and the output is 1 if and only if the solution is feasible. The ESOP expression $F(x_1,\ldots,x_n)$ represents the boolean function $f(x_1,\ldots,x_n)$ in a specific form involving sums of multiple products that are combined through {\small XOR}  operations. In particular, the ESOP expression can be concisely expressed as
\begin{equation}\label{ESOP_definition}
\begin{aligned}
F(x_1,\ldots,x_n) = c_1\oplus c_2 \oplus \ldots \oplus c_m,
\end{aligned}
\end{equation}
where each $c_i$ represents a cube constituted by the product of variables obtained via the {\small AND} operation, and $\oplus$ stands for the modulo-2 addition.

We first recall some notations outlined in \cite{Stergiou2004}. Let $x_i^j$ represent $\overline{x_i}$, $x_i$ and $1$ for $j = 0,\ 1$ and $2$, respectively. For instance, a cube $c_i$ of the form $c_i = x_1^1 \land x_2^0 \land x_3^2$ corresponds to the expression $x_1 \land \overline{x_2} \land 1$, which is equivalent to $x_1 \land \overline{x_2}$ since $1$ is the identity element for logical conjunction.
For a boolean function \(f(x_1, \ldots, x_n)\), the two subfunctions with respect to a variable \(x_i\), denoted \(f_{i}^{j}\) for \(j \in \{0,1\}\), are defined as
\begin{align*}
f^{0}_{i} &= f(x_1, \ldots, x_{i-1}, 0, x_{i+1}, \ldots, x_n), \\
f^{1}_{i} &= f(x_1, \ldots, x_{i-1}, 1, x_{i+1}, \ldots, x_n),
\end{align*}
where \(i = 1, \ldots, n\).
These subfunctions allow us to analyze the behavior of $f$ with respect to each input variable $x_i$ independently.

The initial ESOP expression for a boolean function $f(x_1,\ldots,x_n)$ can be obtained by employing the Shannon expansion \cite{Stergiou2004},
\begin{equation}\label{shannon_expansion}
f(x_1,\ldots,x_n) = x_i^0 \land f_i^0 \oplus x_i^1 \land f_i^1,
\end{equation}
where $x_i^j \land f_i^j$ represents the logical {\small AND} operation between $x_i^j$ and the subfunction $f_i^j$. From the Shannon expansion (\ref{shannon_expansion}), it is evident that each subfunction is itself a boolean function, involving fewer variables than the original function $f$. Consequently, by recursively applying the Shannon expansion to these subfunctions, we can systematically derive the ESOP expressions from the original boolean function $f(x_1,\ldots,x_n)$.

To clarify, consider the ESOP expression for the boolean function,
$ f(x_1,x_2,x_3,x_4) = (x_1 \vee x_2) \wedge (x_2 \vee x_3)$.
Using the Shannon expansion, we partition $f$ into
$ f(x_1,x_2,x_3,x_4) = x_1^0 \land f_1^0 \oplus x_1^1 \land f_1^1$,
where $f_1^0 = x_2$ when $x_1 = 0$, and $f_1^1 = x_2 \vee x_3$ when $x_1 = 1$.
Next, we further decompose $f_1^1$ to
$ f_1^1 = x_2^0 \land x_3 \oplus x_2^1$. Altogether we have the ESOP expression
$ f(x_1,x_2,x_3,x_4) = (x_1^0 \land x_2) \oplus [x_1^1 \land (x_2^0 \land x_3 \oplus x_2^1)]$,
which further reduces to
$ f(x_1,x_2,x_3,x_4) = \overline{x_1} \land x_2 \oplus x_1 \land \overline{x_2} \land x_3 \oplus x_1 \land x_2$. This is a valid ESOP expression for the boolean function $f(x_1,x_2,x_3,x_4)$.

\begin{figure}
\begin{centering}
\includegraphics[scale=0.6]{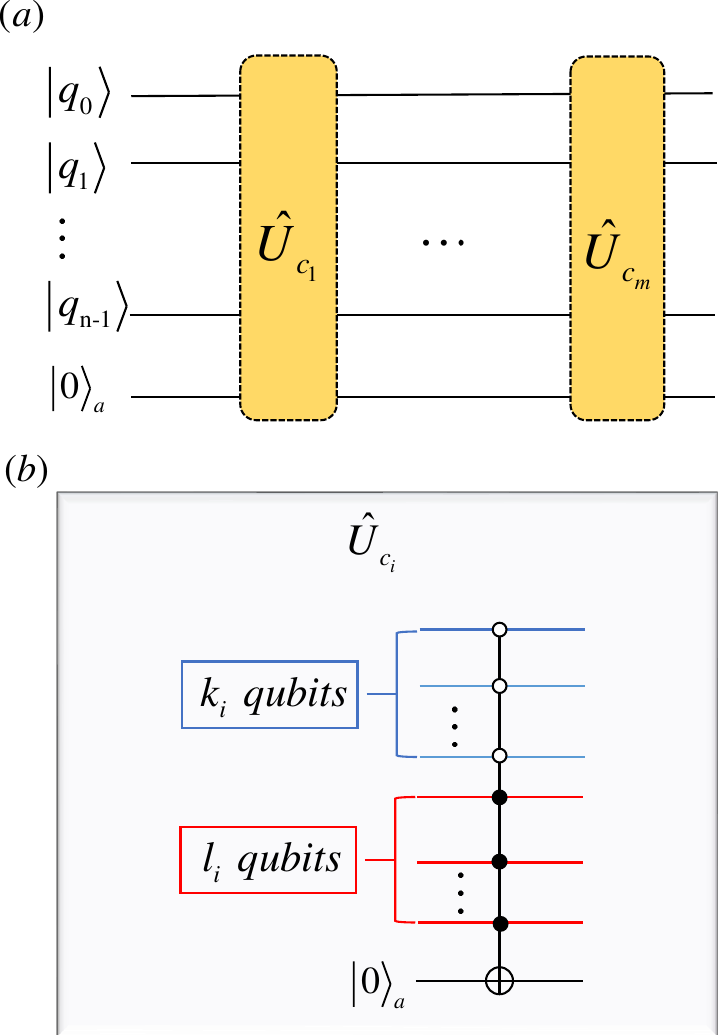}\\
\caption {Quantum circuit designed to implement the oracle $\hat{U}_v$ from the ESOP expression $F=c_1\oplus c_2 \oplus \ldots \oplus c_m$. (a) Global quantum circuit realization. (b) Circuit realization for the operator $\hat{U}_{c_i}$ involved in (a). The $k_i$ qubits (marked by blue lines) denote the negated variables of the cube $c_i$ and $l_i$ qubits (marked by red lines) embody the positive variables of the cube $c_i$.}
\label{fig:valid_circiut}
\end{centering}
\end{figure}

After discussing the representational capability of the ESOP formulation, we next analyze its implementation efficiency for constructing the validation oracle $\hat{U}_v$. This efficiency depends on two factors: the minimized cube count \(m\) and the quantum resource cost per cube (Fig. \ref{fig:valid_circiut}).

While for NPO problems, the polynomial-time verifiability of solution feasibility theoretically guarantees the existence of a cube count \(m\) that scales polynomially, the initial ESOP formulation derived from Shannon decomposition may be rather complex. Therefore, heuristic minimization techniques \cite{Stergiou2004,Mishchenko2001} are employed to obtain a compact, practical expression. Furthermore, the efficiency of the final circuit further hinges on implementing the multi-qubit Toffoli gates (up to $n$-controls) required for each cube. Recent advances in quantum circuit synthesis have yielded highly optimized decompositions for such gates \cite{Mastorakis2025,Claudon2024,Nie2024,Silva2025}. For example, \cite{Silva2025} presents an notable scheme to implement an $n$-qubit Toffoli gate using only a single clean ancilla qubit with a total of  \(6n + 2\) controlled-{\small NOT} ({\small NOT}) gates and depth \(O(\log n)\).

By integrating a minimized ESOP representation with a polynomial number of terms (\(m\)) and these efficient multi-controlled gate implementations, the ESOP-based approach for constructing oracle $\hat{U}_v$ for NPO problems is of practical effectiveness.

\section{VARIATIONAL QUANTUM ALGORITHMS FOR CONSTRAINED COMBINATORIAL OPTIMIZATION PROBLEMS}
\label{sec:VQAs}
Before presenting our method, we first review the algorithmic frameworks of penalty-based methods and ansatz-based schemes, focusing on their principal limitations: the performance bottlenecks of the former and the high circuit complexity of the latter.

Penalty-based methods encode constrained optimization problems into Ising Hamiltonians by adding penalty terms to the problem Hamiltonian $\hat{C}$ for constraint violations:
\begin{equation}\label{penalty-based}
\hat{C} = \hat{\mathcal{O}} + \lambda \hat{S},
\end{equation}
where $\hat{\mathcal{O}}$ encodes the objective function, $\hat{S}$ encodes constraints, and $\lambda > 0$ is the penalty factor. Theoretically, $\lambda$ must exceed a threshold to ensure that the ground state of $\hat{C}$ corresponds to the optimal feasible solution \cite{lihuimin2023Ising,Lucas2014}.

Therefore, penalty-based methods aim to minimize the cost function \( F_p(\vec{\gamma},\vec{\beta}) = \langle \hat{C} \rangle = \langle \hat{\mathcal{O}} \rangle + \lambda \langle \hat{S} \rangle \) under the wave-function ansatz $\ket{\vec{\gamma},\vec{\beta}}$. The performance of this approach is governed by a critical trade-off in the setting of the penalty factor \(\lambda\).
\begin{enumerate}

    \item \textbf{Large \(\lambda\)} causes the penalty term \(\lambda \langle \hat{S} \rangle\) to dominate \(F_p\). This forces the optimization to over-prioritize constraint satisfaction, thereby weakening the optimization of the original objective \(\langle \hat{\mathcal{O}} \rangle\) and hindering the discovery of high-quality solutions .

    \item \textbf{Small \(\lambda\)}, when set to the minimum value theoretically ensuring the ground state corresponds to a feasible solution, results in a solution landscape where feasible states are surrounded by a vast number of infeasible ones, significantly complicating the search process.

\end{enumerate}

Furthermore, the objective term \(\langle \hat{\mathcal{O}} \rangle\) is only meaningful within the feasible region. Consequently, any effort the optimizer expends searching the infeasible region is ineffective and even misleading for improving the true objective .

These inherent contradictions create inefficient and misleading optimization dynamics: the algorithm requires excessive resources to converge to the feasible subspace, severely limiting its scalability and practical performance.

The ansatz-based approach employs a distinct strategy: It encodes constraints directly into the mixing operator \(\hat{B}\) rather than adding penalty terms to the Hamiltonian \(\hat{C}\). This design ensures that the mixing operation \(e^{-i\beta \hat{B}}\) is confined to the feasible subspace, theoretically guaranteeing that all output solutions satisfy the constraints. However, implementing this constrained mixing operation typically requires a complex quantum circuit, often involving multiple calls to a validation oracle \(\hat{U}_v\) and the introduction of large additional ancilla qubits. This significantly increases the circuit depth and overall complexity. For instance, one scheme presented in \cite{Marsh2019} requires \(O(\text{poly}(n) p m)\) calls to the validation oracle \(\hat{U}_v\), where \(m = O(n\beta^2/\epsilon)\), \(\epsilon\) is the simulation precision, \(p\) is the QAOA circuit depth, $\beta$ is the angle parameter in the wave-function ansatz, and \(n\) is the problem size. This scheme also necessitates \(n+3\) ancilla qubits .

 Furthermore, the ansatz-based method requires the preparation of an initial state that is already feasible. While preparing such a state is relatively straightforward for some NPO problems, it also relies on the validation oracle \(\hat{U}_v\) for general NPO problems. Deep quantum circuits (from frequent oracle calls) and a high ancilla-qubit demand hinder the practical realization of these strictly constrained ansatz methods on noisy intermediate-scale quantum devices.

In contrast, to overcome the inefficient sampling of penalty-based methods and the implementation barriers of ansatz-based mixers, we introduce an alternative VQA. Instead of mapping the original problem to a single Hamiltonian, we construct a different type of loss function by using the validation oracle only once.  Our loss function guides the optimizer effectively by creating distinct pathways for feasible and infeasible regions. This simultaneously resolves the decaying probability of effective sampling as the problem size grows in penalty-based methods and avoids the heavy circuit overhead associated with ansatz-based approaches.

We now detail our method by introducing its core components: the wave-function ansatz and the an alternative design of the loss function.

\subsection{Wave-function ansatz with feasibility certification}
\label{sec:wave-function ansatz}
The wave-function ansatz is a cornerstone of VQAs. It critically influences performance by defining the structure of the parametrized quantum circuit and the strategy for optimizing the parameters \( \vec{\theta} \). To efficiently address constrained combinatorial optimization problems, we propose a specialized ansatz that inherently identifies feasible solutions, thereby satisfying problem constraints through its structural design. The ansatz is formulated as
\begin{equation}\label{wave_function_ansatz}
\begin{aligned}
\ket{\Psi(\vec{\theta})}&=\hat{U}_{v}|0\rangle_a|\psi(\vec{\theta})\rangle\\
&= \hat{U}_{v}|0\rangle_a\hat{U}(\vec{\theta})|0\rangle^{\otimes N}\\
&=\sqrt{p_0(\vec{\theta})}\ket{0}_{a}\ket{\psi_{nf}(\vec{\theta})}+
\sqrt{p_1(\vec{\theta})}\ket{1}_{a}\ket{\psi_{f}(\vec{\theta})}.
\end{aligned}
\end{equation}
In this framework, \(N\) denotes the number of qubits required to encode the problem’s solutions and \(\hat{U}(\vec{\theta})\) represents a parametrized variational quantum circuit that constitutes the core of the ansatz. The ancilla qubit \(\ket{0}_a\) serves as a feasibility certification mechanism, with its associated probability amplitudes \(p_0(\vec{\theta})\) and \(p_1(\vec{\theta})\) constrained by \(p_0(\vec{\theta}) + p_1(\vec{\theta}) = 1\). The quantum states \(\ket{\psi_{nf}(\vec{\theta})}\) and \(\ket{\psi_{f}(\vec{\theta})}\) encode infeasible and feasible solutions, respectively. The validation operator $\hat{U}_{v}$ governs the transformation of the combined state \(\ket{0}_a\ket{\psi(\vec{\theta})}\) into a superposition of feasible and infeasible components
$$
\hat{U}_{v}: \ket{0}_a\ket{\psi(\vec{\theta})} \mapsto \sqrt{p_0(\vec{\theta})}\ket{0}_a\ket{\psi_{nf}} + \sqrt{p_1(\vec{\theta})}\ket{1}_a\ket{\psi_{f}}.
$$

In this work, we construct the unitary operator $\hat{U}(\vec{\theta})$ by simultaneously considering both optimization objective and constraint of the constrained combinatorial optimization problem. We define the problem using two Hamiltonians: $\hat{\mathcal{O}}=\sum_i \hat{o}_i$ for the objective and $\hat{S}=\sum_i \hat{s}_i$ for the constraint, following the formalism in \cite{lihuimin2023Ising}. Then the unitary operator is given by
\begin{equation}
\begin{aligned}\label{unitary_operator}
\hat{U}(\vec{\theta})= \hat{U}_{M}(\vec{\beta_{p}})\hat{U}_{D}(\vec{\gamma_{p}},\vec{\mu_p}),\ldots,\hat{U}_{M}(\vec{\beta_{1}}) \hat{U}_{D}(\vec{\gamma_1},\vec{\mu_1})\hat{H}^{\otimes N},
\end{aligned}
\end{equation}
with
\begin{equation}\label{prob_operator}
\begin{aligned}
\hat{U}_D(\vec{\gamma_{l}},\vec{\mu_{l}})&:=e^{(-i\sum_{j}\mu^{(l)}_{j}\hat{s}_j)}e^{(-i\sum_{j}\gamma^{(l)}_{j}\hat{o}_j)},\\
\hat{U}_M(\vec{\beta_{l}})&:=e^{(-i\sum_{j=0}^{j=N-1}\beta^{(l)}_{j}X_j)},
\end{aligned}
\end{equation}
where $\vec{\gamma_l}=(\ldots,\gamma^{(l)}_{j},\ldots)$, $\vec{\mu_l}=(\ldots,\mu^{(l)}_{j},\ldots)$, $\vec{\beta_l}=(\ldots,\beta^{(l)}_{j},\ldots)$, $\vec{\theta}=(\vec{\beta_{1}},\vec{\gamma_{1}},\vec{\mu_1},\ldots, \vec{\beta_{p}}, \vec{\gamma_{p}},\vec{\mu_p})$, $\hat{X}_j$ denotes the standard Pauli operator $\hat{X}$ on the $j$th spin, $\hat{H}$ is the Hadamard operator, and $p$ is the depth of the variational circuit.

\subsection{Feasibility-guiding loss function}
\label{sec:loss_function}

$Theorem\ \emph{1}.$ Let \(\hat{\mathcal{O}}\) and \(\hat{S}\) denote the Ising Hamiltonians encoding the optimization objective and the constraints, respectively, for a given constrained combinatorial optimization problem. The optimal feasible solution can be obtained through minimizing the loss function
\begin{widetext}
\begin{equation}\label{cost_function}
\begin{aligned}
E(\vec{\theta})&=\langle \Psi(\vec{\theta})|\big(|1\rangle_a\langle 1|_a\otimes (\hat{\mathcal{O}}-E_{\mathcal{O}}I)+|0\rangle_a\langle 0|_a\otimes (\hat{S}-E_{S}I)\big)|\Psi(\vec{\theta})\rangle,
\end{aligned}
\end{equation}
\end{widetext}
where $E_{\mathcal{O}}$ and $E_{S}$ denote the maximum eigenvalue of $\hat{\mathcal{O}}$ and the minimum eigenvalue of $\hat{S}$, respectively.

$Proof.$ Substituting $\ket{\Psi(\vec{\theta})}$ \eqref{wave_function_ansatz} into the cost function $E(\vec{\theta})$, we have
\begin{widetext}
\begin{equation}
\begin{aligned}
E(\vec{\theta})&=\langle \Psi(\vec{\theta})|\big(|1\rangle_a\langle 1|_a\otimes (\hat{\mathcal{O}}-E_{\mathcal{O}}I)+|0\rangle_a\langle 0|_a\otimes (\hat{S}-E_{S}I)\big)|\Psi(\vec{\theta})\rangle\\
&=(\sqrt{p_0(\vec{\theta})}\langle0|_a\langle\psi_{nf}(\vec{\theta})|
+\sqrt{p_1(\vec{\theta})}\langle1|_a\langle\psi_f(\vec{\theta})|)|1\rangle_a\langle 1|_a\otimes (\hat{\mathcal{O}}-E_{\mathcal{O}}I)(\sqrt{p_0(\vec{\theta})}\ket{0}_{a}\ket{\psi_{nf}(\vec{\theta})}+
\sqrt{p_1(\vec{\theta})}\ket{1}_{a}\ket{\psi_{f}(\vec{\theta})})\\
&+(\sqrt{p_0(\vec{\theta})}\langle0|_a\langle\psi_{nf}(\vec{\theta})|
+\sqrt{p_1(\vec{\theta})}\langle1|_a\langle\psi_f(\vec{\theta})|)|0\rangle_a\langle 0|_a\otimes (\hat{S}-E_{S}I)(\sqrt{p_0(\vec{\theta})}\ket{0}_{a}\ket{\psi_{nf}(\vec{\theta})}+
\sqrt{p_1(\vec{\theta})}\ket{1}_{a}\ket{\psi_{f}(\vec{\theta})})\\
&=p_1(\vec{\theta})\langle\psi_f(\vec{\theta})|(\hat{\mathcal{O}}-E_{\mathcal{O}}I)\ket{\psi_{f}(\vec{\theta})}+p_0(\vec{\theta})\langle\psi_{nf}(\vec{\theta})|(\hat{S}-E_{S}I)\ket{\psi_{nf}(\vec{\theta})}.
\end{aligned}
\end{equation}
\end{widetext}
Based on the choice of $E_\mathcal{O}$ and $E_S$, it follows that
\begin{equation}
\begin{aligned}
\langle\psi_f(\vec{\theta})|(\hat{\mathcal{O}}-E_\mathcal{O} I)|\psi_f(\vec{\theta})\rangle \le 0, \\
\langle\psi_{nf}(\vec{\theta})|(\hat{S}-E_S I)|\psi_{nf}(\vec{\theta})\rangle \ge 0,
\end{aligned}
\end{equation}
and since $\hat{\mathcal{O}}-E_\mathcal{O} I$ preserves the ordering of the eigenvalues of $\hat{\mathcal{O}}$, the loss function $E(\vec{\theta})$ is minimized at $\vec{\theta}^*$ when $p_1(\vec{\theta}^*)=1$ and $|\Psi(\vec{\theta}^*)\rangle = |1\rangle_a |\psi_f(\vec{\theta}^*)\rangle$, where $|\psi_f(\vec{\theta}^*)\rangle$ encodes the optimal feasible solution to the original constrained combinatorial optimization problem. $\Box$

Based on Theorem 1, our loss function ensures universally higher values for infeasible solutions while establishing distinct computational pathways for feasible versus infeasible regions during iterative optimization. This provides clear and non-conflicting guidance to the optimizer, thereby avoiding the extensive inefficient and even misleading sampling in infeasible regions characteristic of penalty-based methods during early iterations. Furthermore, it mitigates the complex optimization landscape arising from the conflicting objectives between the target function and constraints in penalty-based approaches. As a result, the algorithm’s overall performance is significantly enhanced.

To bridge these theoretical advantages with practical implementation, we note that for applying Theorem 1, the exact maximum eigenvalue of $\hat{\mathcal{O}}$ and the exact minimum eigenvalue of $\hat{S}$ need not be known precisely. It suffices to set $E_{\mathcal{O}}$ and $E_{S}$ to a known upper bound and a known lower bound for these values, respectively. To clearly show how to estimate the cost function under this practical consideration, we rewrite $E(\vec{\theta})$ as

\begin{widetext}
\begin{equation}
\begin{aligned}\label{loss_function_CCOP}
E(\vec{\theta}) &= \langle \Psi(\vec{\theta})|\big(|1\rangle_a\langle 1|_a\otimes (\hat{\mathcal{O}}-E_{\mathcal{O}}I)\big)|\Psi(\vec{\theta})\rangle + \langle \Psi(\vec{\theta})|\big(|0\rangle_a\langle 0|_a\otimes (\hat{S}-E_{S}I)\big)|\Psi(\vec{\theta})\rangle\\
&=\frac{1}{2}\langle \Psi(\vec{\theta})|(I_a-\hat{Z}_{a})\otimes (\hat{\mathcal{O}}-E_{\mathcal{O}}I)|\Psi(\vec{\theta})\rangle+\frac{1}{2}\langle \Psi(\vec{\theta})|(I_a+\hat{Z}_{a})\otimes (\hat{S}-E_{S}I)|\Psi(\vec{\theta})\rangle\\
&=\frac{1}{2}\langle \Psi(\vec{\theta})|\big(I_a\otimes \hat{\mathcal{O}}-\hat{Z}_{a}\otimes \hat{\mathcal{O}}+E_{\mathcal{O}}\hat{Z}_{a}\big)|\Psi(\vec{\theta})\rangle+\frac{1}{2}\langle \Psi(\vec{\theta})|\big(I_a\otimes \hat{S}+\hat{Z}_{a}\otimes \hat{S}-E_{S}\hat{Z}_{a}\big)|\Psi(\vec{\theta})\rangle-\frac{1}{2}(E_{\mathcal{O}}+E_{S})\\
&=\frac{1}{2}\langle \Psi(\vec{\theta})|\big(I_a\otimes \hat{\mathcal{O}}-\hat{Z}_{a}\otimes \hat{\mathcal{O}}+I_a\otimes \hat{S}+\hat{Z}_{a}\otimes \hat{S}+(E _{\mathcal{O}}-E_{S})\hat{Z}_{a}\big)|\Psi(\vec{\theta})\rangle-\frac{1}{2}(E_{\mathcal{O}}+E_{S}),
\end{aligned}
\end{equation}
\end{widetext}
where $\hat{Z}_{a}$ is the standard Pauli $\hat{Z}$ operator applied to the auxiliary qubit.

\section{NUMERICAL EXPERIMENTS}
\label{sec:Experiments}
This section presents a numerical evaluation of our proposed algorithm against penalty-based methods for the minimum-vertex-cover (MVC) and maximum-independent-set (MIS) problems, to illustrate the limitations of the latter and underscore the advantages of the former. A direct comparison with ansatz-based schemes is omitted due to their substantial implementation overhead.

Specifically, we evaluate our algorithm's advantages by focusing on two aspects: analyzing the impact of penalty factors on the performance of penalty-based methods, and comparing the capability of both algorithms to escape local optima under diverse initializations. Within the VQA framework, the cost function landscape is typically complex and riddled with local minima. Conducting optimizations from multiple random starting points increases the probability that at least one initial point lies within the basin of attraction of the global optimum, thereby enhancing the likelihood of escaping local minima. By performing independent optimizations from multiple random initial parameter sets and comparing the outcomes, we assess this escape capability.

\subsection{Analysis of minimum-vertex-cover cases}
\label{subsection:mvc}

Given an undirected graph $G=(V,E)$ with $V$ the set of vertices and $E$ the set of edges, a vertex set $S \subseteq V$ covers an edge if at least one of the edge's endpoints is in $S$. The MVC problem seeks to find the smallest vertex set $S$ that covers all edges in $E$. In other words, the optimization objective is to minimize the number of vertices in the set $S$ and the constraint is to ensure all the edges of graph $G$ are covered by the set $S$.

Let the binary bit $z_i$ represent the state of the $i$th vertex in graph $G$, where $z_i = -1$ (spin down) indicates that the vertex is included in the vertex set $S$ and $z_i = +1$ (spin up) indicates that it is not included in $S$. To quantify the target optimization objective, we employ the function $C_v(\mathbf{z}) = \sum_{i} \frac{1}{2}(1 - z_i)$, which counts the number of vertices in the set $S$ by following the idea introduced in \cite{thomas_monte_2014}. The corresponding Ising Hamiltonian has the form
\begin{equation}\label{eq:H_a_mvc}
\hat{\mathcal{O}}=\sum_{i}(1-\hat{Z}_i)/2,
\end{equation}
where $\hat{Z}_i$ denotes the standard Pauli operator $\hat{Z}$ on the $i$th spin. With respect to the constraint, the corresponding Ising Hamiltonian can be written as \cite{lihuimin2023Ising}
\begin{equation}\label{eq:H_b_mvc}
\hat{S}={\underset {<u,v>\in \text{edges}}{\operatorname{\sum}}} \left(\hat{Z}_u\hat{Z}_v+\hat{Z}_u+\hat{Z}_v+I\right)/4.
\end{equation}

To tackle the MVC problem we initially integrate the Ising Hamiltonians $\hat{\mathcal{O}}$ and $\hat{S}$, then employ the parametrized variational quantum circuit $\hat{U}(\vec{\gamma_{l}},\vec{\mu_{l}})$ as
\begin{equation}\label{prob_operator_mvc}
\begin{aligned}
\hat{U}_D(\vec{\gamma_{l}},\vec{\mu_{l}})&
:=e^{(-i\sum_{j}\mu^{(l)}_{j}\hat{Z}_j)}e^{(-i\sum_{<u,v>}\gamma^{(l)}_{<u,v>}
\hat{Z}_u\hat{Z}_v)}.
\end{aligned}
\end{equation}

 To ensure a fair comparison within the ansatz search space, we independently assign variational parameters to each term of the mixing operator \(\hat{B}\) and the problem operator \(\hat{C}\) (which includes the penalty factor) in the wave-function ansatz of the penalty-based method. This guarantees that the total number of optimizable parameters is consistent between our algorithm and the penalty-based one at the same circuit depth. Consequently, any observed performance difference can be attributed to the algorithmic mechanisms themselves rather than the parameter count.

The problem Hamiltonian for the penalty-based method applied to the MVC problem is given by
\[
\hat{C} = \sum_{\langle i,j \rangle} \frac{\lambda}{4} \hat{Z}_i \hat{Z}_j + \sum_{i=0}^{n-1} \left(-\frac{1}{2} + \frac{\lambda}{4}d_i\right) \hat{Z}_i,
 \]
where \(d_i\) denotes the degree of vertex \(i\). The corresponding wave-function ansatz is constructed as
 \[
\ket{\vec{\gamma}, \vec{\beta}, \vec{\mu}} = \hat{U}(\vec{\theta}) |0\rangle^{\otimes N},
\]
 where
\[
\hat{U}(\vec{\theta}) = \hat{U}_{M}(\vec{\beta}_{p}) \hat{U}_{C}(\vec{\gamma}_{p}, \vec{\mu}_p) \cdots \hat{U}_{M}(\vec{\beta}_{1}) \hat{U}_{C}(\vec{\gamma}_1, \vec{\mu}_1) H^{\otimes N},
\]
with \(\hat{U}_{M}(\vec{\beta}_{l})\) given in \eqref{prob_operator}. Specifically, the unitary operator corresponding to the problem Hamiltonian is defined as
 \[
\hat{U}_{C}(\vec{\gamma}_l, \vec{\mu}_l) := e^{-i \sum_{j} \mu^{(l)}_{j} \left(-\frac{1}{2} + \frac{\lambda}{4}d_j\right) \hat{Z}_j} \; e^{-i \sum_{\langle u,v \rangle} \gamma^{(l)}_{\langle u,v \rangle} \frac{\lambda}{4}\hat{Z}_u \hat{Z}_v}.
 \]

For the construction of the validation oracle \(\hat{U}_v\) in our algorithm, let \(|q_i\rangle\) denote the eigenvector of the Pauli operator \(\hat{Z}_i\) with eigenvalue \(q_i\). The constraint for the MVC problem can be formulated as the boolean function
\begin{equation}\label{eq:H_b_mvc}
F(q_0, \ldots, q_{N-1}) = \bigwedge_{(u,v) \in \text{edges}} (q_u \vee q_v),
\end{equation}
where $q_i$ signifies the binary variable (e.g., 0 or 1) associated with the vertex $i$ in the graph. A feasible solution to the MVC problem is encoded to the quantum state $|q_0, \ldots, q_{N-1}\rangle$, whereas $F(q_0, \ldots, q_{N-1})$ is evaluated to be 1, ensuring that every edge in the graph is covered by at least one of its endpoints being selected.

From the formula~\eqref{eq:H_b_mvc}, we can obtain a compact, practical ESOP for the constraints by applying Shannon decomposition and heuristic minimization techniques \cite{Stergiou2004,Mishchenko2001}. Subsequently, a quantum circuit for the operator $\hat{U}_v$ can be constructed based on this ESOP, as illustrated in Fig.~\ref{fig:valid_circiut}.

To demonstrate the limitations of penalty-based methods, we adopt the following evaluation scheme: Five different penalty factors are uniformly sampled from an interval \((a, b]\), where the lower bound \(a\) is the theoretical threshold ensuring that the Hamiltonian's ground state corresponds to the optimal feasible solution and the upper bound \(b\) is set to \(a + 10\) to avoid excessively large penalties. For MVC problems, $a$ is equal to 1 \cite{lihuimin2023Ising}. Then we evaluate the effect of penalty factors on solution quality by two metrics: average performance (mean accuracy) and performance variation (variance).

 Aiming to ensure a comprehensive and unbiased evaluation, we conduct extensive repeated experiments. For each problem size \(n\), ten MVC instances are randomly generated from the Erd\"os-R\'enyi ensemble with an edge probability of 0.5. For each instance, six sets of initial parameters are generated randomly. The final performance is reported as the statistical average over all random configurations.

Specifically, let \(A_{pij}^{n}\) denote the accuracy achieved for a given penalty factor \(p\), problem instance \(i\), and initial state \(j\) at problem size \(n\). The average performance corresponding to a specific penalty factor \(p\) is calculated by aggregating over all instances and initial states:
\[
\mu_{p}^{n} = \frac{1}{60} \sum_{i=1}^{10} \sum_{j=1}^{6} A_{pij}^{n}.
\]

Subsequently, the overall mean performance \(\mu^n\) and the performance variance \(\sigma^n\) across all 5 penalty factors are computed as follows:

\[
\mu^{n} = \frac{1}{5} \sum_{p=1}^{5} \mu_{p}^{n}, \quad \sigma^{n} = \frac{1}{5} \sum_{p=1}^{5} \left( \mu_{p}^{n} - \mu^{n} \right)^2.
\]

\begin{figure*}[t!]
    \centering
    \includegraphics[scale=1.0]{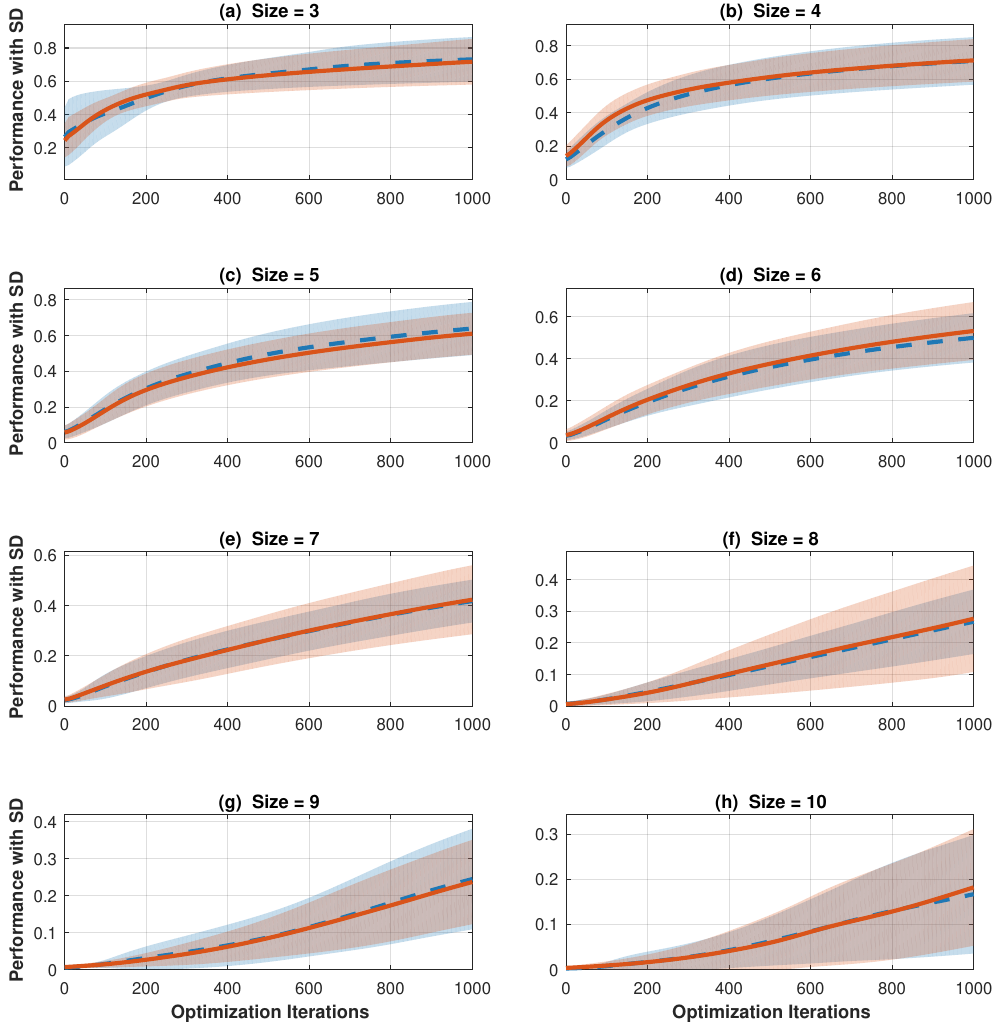}\\
\caption{
Performance of the penalty-based method across problem sizes(a) 3, (b) 4, (c) 5, (d) 6, (e) 7,(f) 8, (g) 9, and (h) 10.
Subfigures (a)--(h) show the results for each size.
Blue dashed lines represent the mean accuracy for depth~2,
while orange solid lines correspond to depth~3.
Blue shaded areas indicate the mean plus or minus one standard deviation for depth~2
and orange shaded areas denote the same for depth~3.
Penalty factors are uniformly sampled from the interval $(a,b]$.
}
    \label{fig:MVC_superparameter}
\end{figure*}

Whether the limiting effect of penalty factors is alleviated with deeper quantum circuits is a key question. We address this by comparing \( \mu^{n} \) and \( \sigma^{n} \) for penalty-based methods at depths 2 and 3 (see Fig.~\ref{fig:MVC_superparameter}).

The results in Fig.~\ref{fig:MVC_superparameter} demonstrate a limited performance of the penalty-based method on MVC problems. Notably, at depth 2, the average accuracy is low even for small-scale instances (e.g., below 0.75 for size 3). More critically, increasing the depth to 3 does not translate into a reliable enhancement in solution quality, as the average accuracy does not show a clear upward trend and the variance does not consistently decrease. This implies that the introduction of a penalty factor may alter the structure of the optimization problem, suggesting that even with an increase in quantum circuit depth, the algorithm struggles to effectively approach the global optimum.

 In contrast to traditional penalty-based methods, our algorithm does not introduce additional hyperparameter penalty factors. In the comparative experiments, to fairly reflect the inherent search efficiency of our algorithms, we design the experiment as follows: The parametrized circuit depth of our proposed algorithm is set to 2, while the depth of the penalty-based method is set to 3. For the penalty-based method, we use a grid search to determine the penalty factor yielding the best average performance and evaluate the influence of initial states based on this. This design is based on the following considerations: Although our depth 2 scheme might involve slightly more {\small CNOT} gates than the depth-3 penalty-based method, the penalty-based method introduces more parameters to be optimized as depth increases, significantly raising the optimization difficulty.  Crucially, as established earlier, the detrimental effect of the penalty factor persists despite the increased depth.

\begin{figure*}[t!]
    \centering
    \includegraphics[scale=1.0]{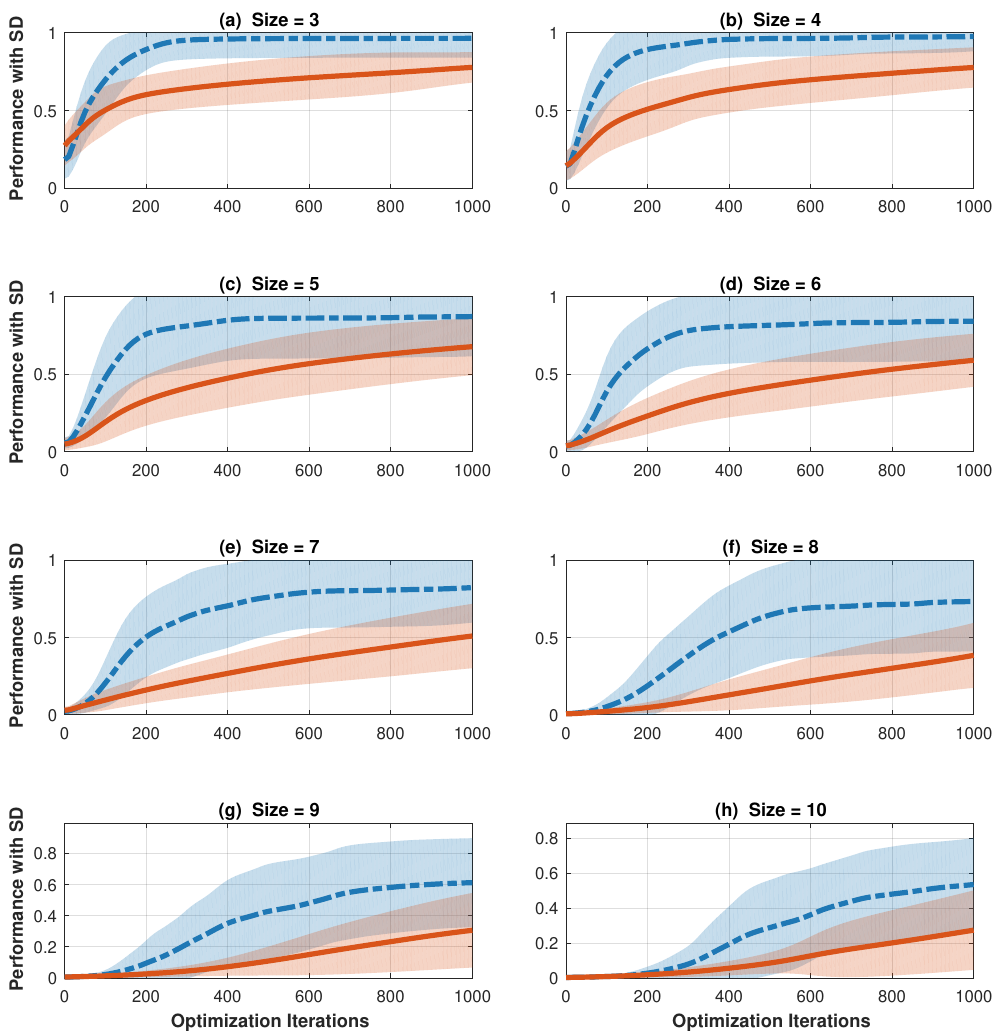}\\

\caption{
Performance comparison between our method and the penalty-based method across problem sizes (a) 3,(b) 4, (c) 5, (d) 6, (e) 7, (f) 8, (g) 9, and (h) 10.
Subfigures (a)--(h) show results for each size, evaluated over six sets of randomly generated initial parameters.
For the penalty-based method, the penalty factor yielding the best average performance among five randomly generated values is selected.
Blue dash-dotted lines represent the mean accuracy of our approach, while orange solid lines denote the penalty-based method.
Shaded regions indicate the mean plus or minus one standard deviation (blue for our method and orange for the penalty-based method).
}
    \label{fig:MVC_initialparameter}
\end{figure*}

\begin{figure*}[t!]
        \centering
        \includegraphics[scale=1.0]{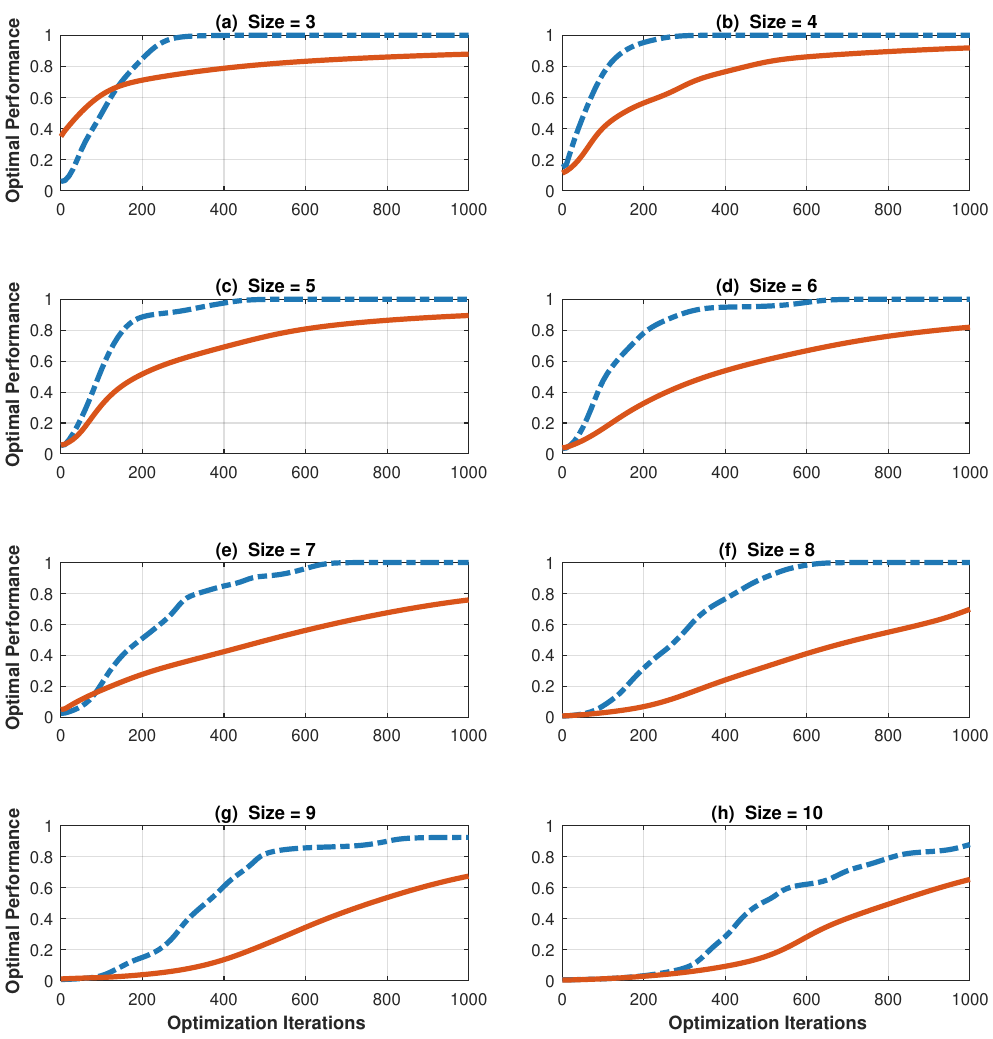}\\
\caption{
Optimal accuracy achieved under the respective optimal initial parameters for each method across different problem sizes (a) 3, (b) 4, (c) 5, (d) 6, (e) 7, (f) 8, (g) 9, and (h) 10. Blue dash-dotted lines are for our approach and orange solid lines for the penalty-based method.
}
    \label{fig:MVC_optimalparameter}
\end{figure*}

 To comprehensively evaluate the performance, we compare both algorithms across six sets of randomly initialized parameters, analyzing both the average performance and performance variance. Furthermore, we directly compare their performance under their respective optimal initialization. This comparison under optimal starting conditions provides a clearer view of each algorithm's peak capability to escape local optima and converge towards high-quality solutions. The results are presented in Figs.~\ref{fig:MVC_initialparameter} and ~\ref{fig:MVC_optimalparameter}.

Figures.~\ref{fig:MVC_initialparameter} and ~\ref{fig:MVC_optimalparameter} show that, even with an optimal penalty factor, the penalty-based method is outperformed by our algorithm in both average and best-case performance across initial states, a gap that widens with problem size. The higher variance of our algorithm for problems larger than size 4 reflects its stronger capacity to escape local optima and converge to high-quality solutions, consistent with its superior accuracy. In contrast, the penalty-based method exhibits not only lower performance but also limited variance, indicating a tendency to become trapped in local minima and a constrained exploration of the solution space.

\subsection{Analysis on maximum-independent-set cases}
\label{subsection:mis}

The MIS problem involves finding an independent set $S \subseteq V$ for a given graph $G = (V, E)$ that contains the largest possible number of vertices, where an independent set is defined as a subset of vertices in which any pair of vertices are not adjacent. In other words, the optimization object for MIS problem is to maximize the number of vertices in $S$ and the related constraint is to ensure no edges between any pair of vertices in the set.

Similar to the MVC problem, the Ising Hamiltonian for the optimization object has the following form \cite{lihuimin2023Ising}
\begin{equation}\label{eq:H_b_mis}
\hat{\mathcal{O}}=nI-\sum_{i=0}^{n-1}(I-\hat{Z}_i)/2.
\end{equation}
The constraint for the MIS problem can be expressed as the boolean function
\begin{equation}\label{eq:H_b_mis}
F(q_0, \ldots, q_{N-1}) = \bigvee_{(u,v) \in \text{edges}} (q_u \land q_v),
\end{equation}
where a valid solution to the MIS problem corresponds to a quantum state $|q_0, \ldots, q_{N-1}\rangle$ which vanished $F(q_0, \ldots, q_{N-1})$. To analyze this constraint in a manner analogous to the MVC case, we consider the complement of \(F\), defining the function \(G\) as
\begin{equation}\label{eq:H_b_mis1}
G(q_0, \ldots, q_{N-1}) = \overline{F}(q_0, \ldots, q_{N-1}) = \bigwedge_{(u,v) \in \text{edges}} (\overline{q_u} \vee \overline{q_v}).
\end{equation}

Consequently, a feasible solution to the MIS problem is encoded in the quantum state $|q_0, \ldots, q_{N-1}\rangle$ so that $G(q_0, \ldots, q_{N-1})$ evaluates to 1. Leveraging this, the quantum circuit for the operator $\hat{U}_v$ can be constructed similar to that of MVC problems, while the parametrized variational quantum circuit $\hat{U}_D(\vec{\gamma_{l}},\vec{\mu_{l}})$ is also adopted in the form given in~\eqref{prob_operator_mvc}.

Using the experimental framework established for MVC, we benchmark our algorithm against the penalty-based method on the MIS problem, analyzing both the impact of penalty factors and the relative capability of each method to escape local optima.

\begin{figure*}[t!]
    \centering
        \includegraphics[scale=1.0]{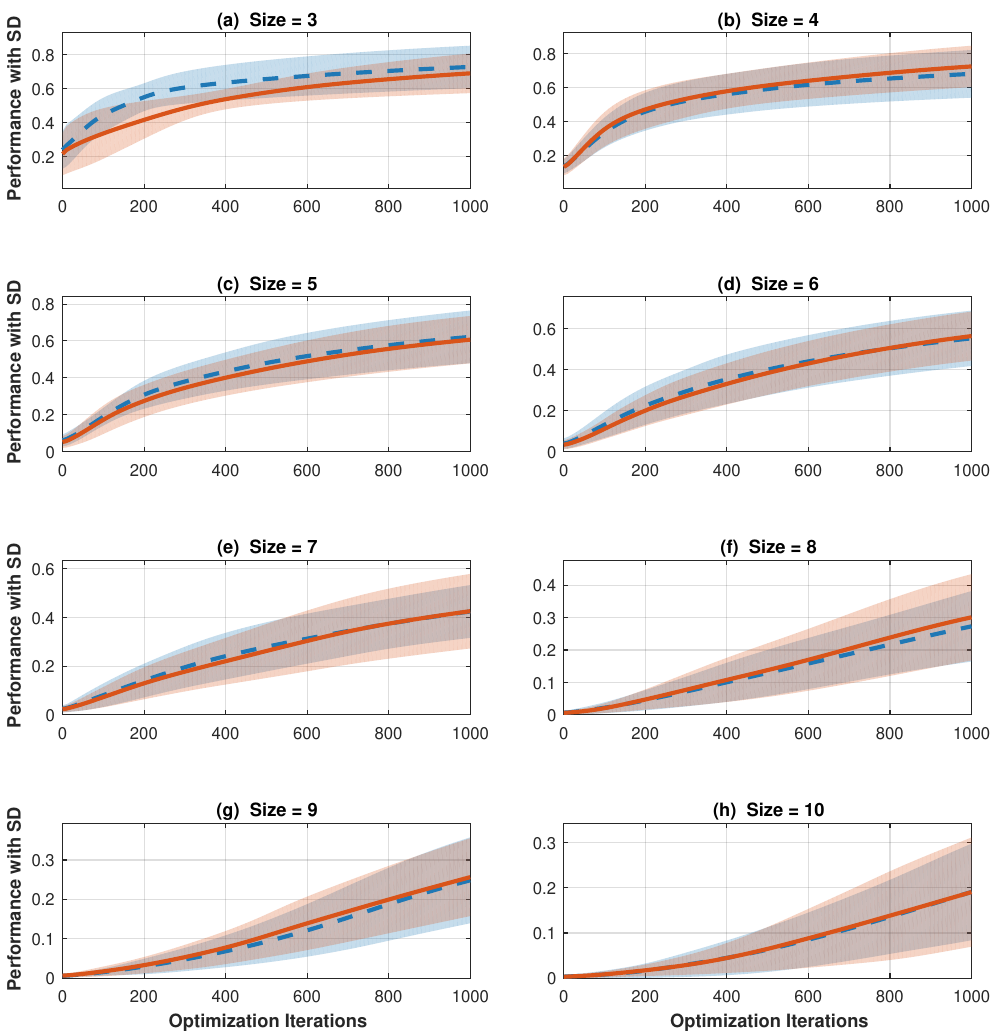}\\
\caption{
Performance of the penalty-based method across different problem sizes (a) 3, (b) 4, (c) 5, (d) 6, (e) 7,(f) 8, (g) 9, and (h) 10.
Blue dashed lines represent the mean accuracy for depth~2,
while orange solid lines correspond to depth~3.
Blue shaded areas indicate the mean plus or minus one standard deviation for depth~2
and orange shaded areas denote the same for depth~3.
Penalty factors are uniformly sampled from the interval $(a,b]$.
}
    \label{fig:MIS_superparameter}
\end{figure*}

\begin{figure*}[t!]
    \centering\includegraphics[scale=1.0]{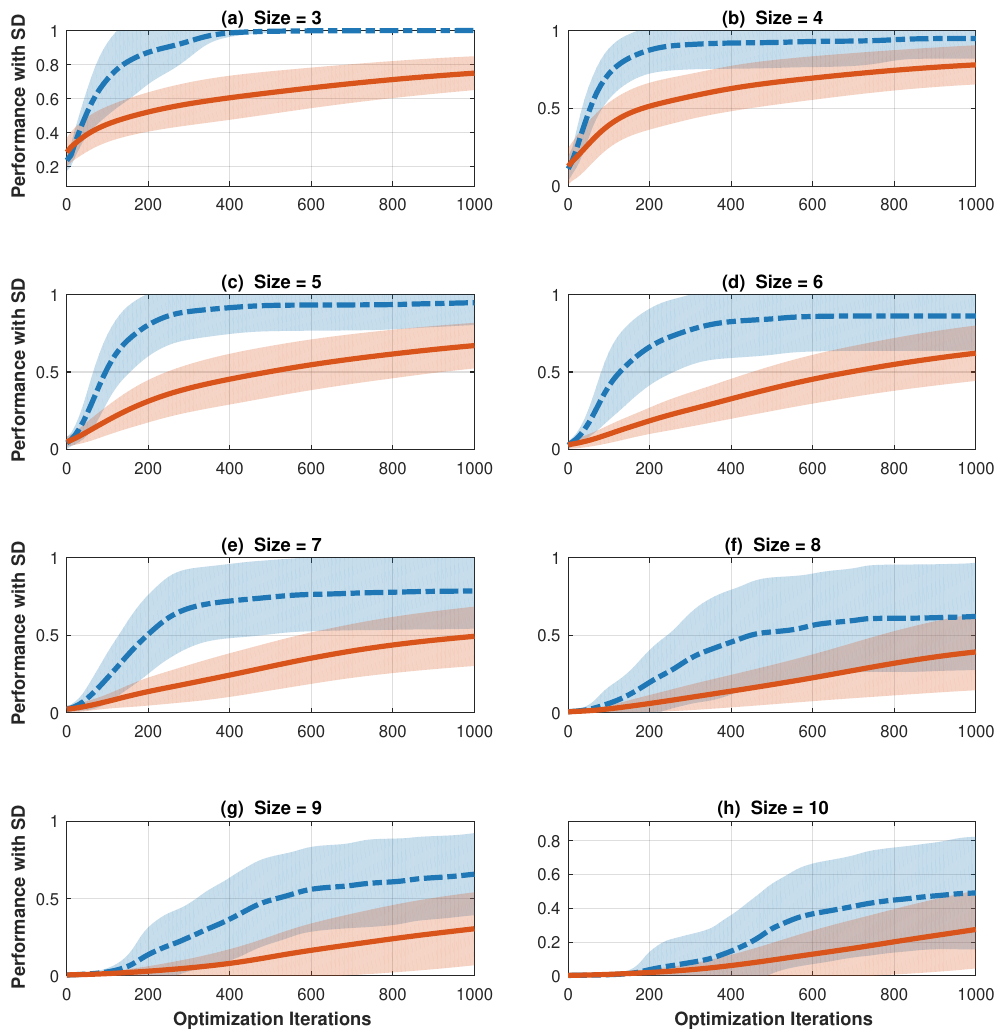}\\
\caption{
Performance comparison between our method and the penalty-based method across problem sizes (a) 3,(b) 4, (c) 5, (d) 6, (e) 7, (f) 8, (g) 9, and (h) 10, evaluated over 6 sets of randomly generated initial parameters.
For the penalty-based method, the penalty factor yielding the best average performance among five randomly generated values is selected.
Blue dash-dotted lines represent the mean accuracy of our approach, while orange solid lines denote the penalty-based method.
Shaded regions indicate the mean plus or minus one standard deviation (blue for our method and orange for the penalty-based method).
}
    \label{fig:MIS_initialparameter}
\end{figure*}

\begin{figure*}[t!]
    \centering
        \centering\includegraphics[scale=1.0]{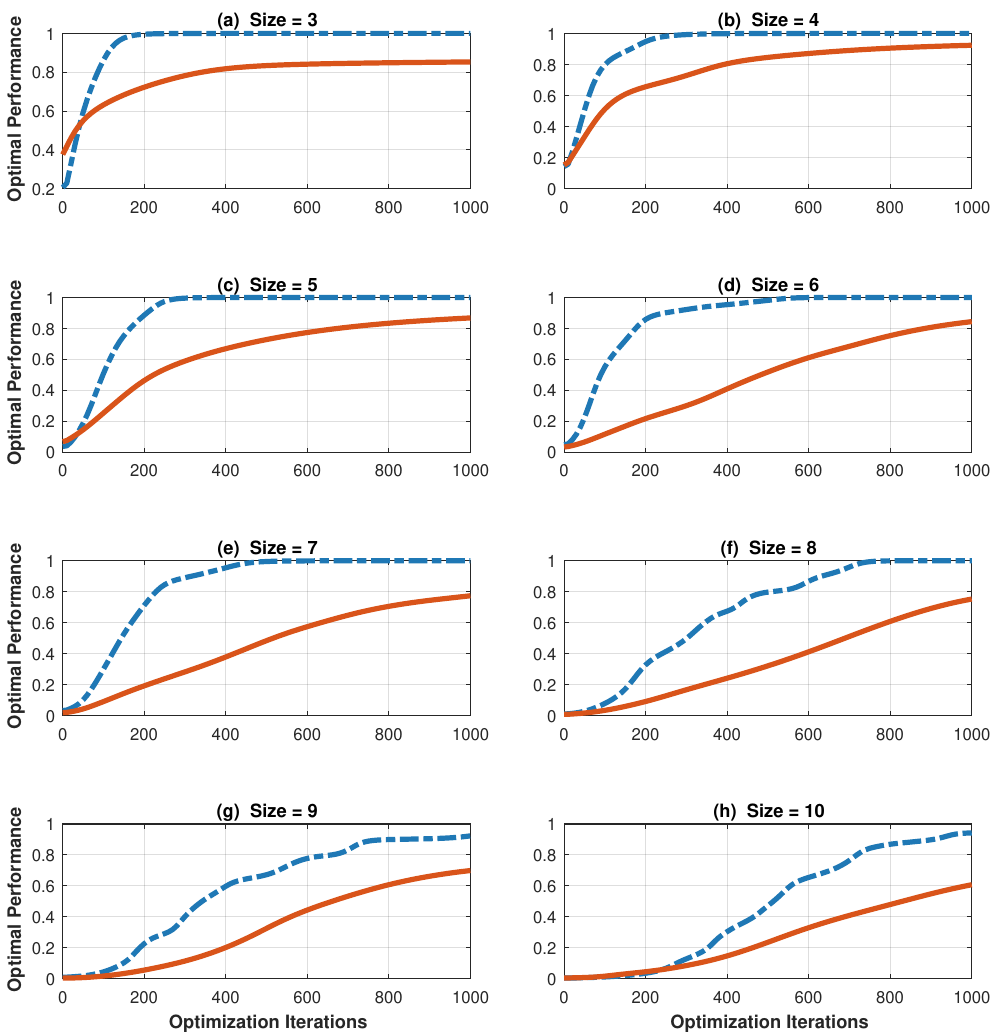}\\
\caption{
Optimal accuracy achieved under the respective optimal initial parameters for each method across different problem sizes (a) 3, (b) 4, (c) 5, (d) 6, (e) 7, (f) 8, (g) 9, and (h) 10. Blue dash-dotted lines for our approach and orange solid lines for the penalty-based method.
}
    \label{fig:MIS_optimalparameter}
\end{figure*}

As shown in Fig.~\ref{fig:MIS_superparameter}, the influence of the penalty factor on algorithm performance when applied to the MIS problem is consistent with its effect on the MVC problem. At a depth of 2, the penalty-based method yields low accuracy (e.g., below 0.75 even for small-scale instances of size 3). Furthermore, increasing the depth to 3 fails to produce a reliable improvement in solution quality and may even degrade performance on certain problem sizes. Similarly, the superior performance of our algorithm in escaping local optima is also consistent across both problems, as demonstrated by the comparative results in Figs.~\ref{fig:MIS_initialparameter} and ~\ref{fig:MIS_optimalparameter}.

\section{Conclusion}
\label{conclusion}
For constrained combinatorial optimization problems, conventional approaches often face a critical trade-off: They either rely on prohibitively complex quantum circuit architectures (ansatz-based approaches) or suffer from inefficient and even leading sampling in vast infeasible regions (penalty-based methods). To overcome this, we have introduced an alternative VQA with a strategically designed loss function.

The design of the VQAs hinges on formulating both a parametrized wave-function ansatz and a cost function whose global minimum corresponds to the optimal feasible solution. We constructed an ansatz capable of feasibility certification using a validation oracle $\hat{U}_v$. On this basis, our loss function was designed to assign consistently higher values to infeasible solutions while creating distinct computational pathways for feasible and infeasible regions during optimization. This provides the optimizer with clear, nonconflicting guidance, thereby preventing the extensive, often inefficient and even misleading sampling in infeasible regions that characterizes penalty-based methods in early iterations. Furthermore, it mitigates the complex optimization landscape caused by conflicting objectives in penalty-based approaches.

Our algorithm focuses on NPO problems, a class of constrained combinatorial optimization problems for which determining the feasibility of a solution can be done efficiently. For such problems, the existence of an efficient validation oracle, denoted by \(\hat{U}_v\), can be proven. This serves as a foundational assumption of our approach. Regarding circuit complexity, our design requires only one additional efficient validation oracle module and one ancilla qubit beyond the penalty-based method. In contrast, ansatz-based methods typically necessitate multiple calls to the validation oracle \(\hat{U}_v\) and a significant number of additional ancilla qubits.

To numerically demonstrate the advantages of our approach, we benchmarked its performance against the penalty-based method on MVC and MIS problems of sizes 3-10. A direct performance comparison with ansatz-based schemes was precluded by their considerable implementation complexity. Specifically, we evaluated our algorithm’s advantages by focusing on two aspects: analyzing the impact of penalty factors on the performance of penalty-based methods, and comparing the capability of both algorithms to escape local optima under diverse initializations.

Based on the numerical analysis of penalty factors across both MVC and MIS problems, the performance of the penalty-based method remained limited. At a depth of 2, the algorithm achieved low accuracy even for small-scale instances (e.g., below 0.75 for problems of size 3). Furthermore, increasing the depth to 3 failed to yield a reliable improvement in solution quality and even degraded performance on certain problem sizes.

In contrast, our algorithm operates without hyperparameter penalty factors. To ensure a fair comparison of inherent efficiency, we evaluated our algorithm at a circuit depth of 2 against the penalty-based method at a depth of 3. The numerical results confirm the superior performance of our method over penalty-based approaches in solving the MVC and MIS problems, demonstrating its enhanced capability to escape local optima and converge to the global optimum. These findings provide numerical evidence for the improved optimization landscape facilitated by our designed loss function.
\\

\begin{acknowledgements} This work was supported by National Natural Science Foundation of China Grant No.12171044, the specific research fund of the Innovation Platform for Academicians of Hainan Province, and the Fundamental Research Funds for the Central Universities.
\end{acknowledgements}

\end{document}